\documentclass[aip, jmp,showpacs,amsmath,amssymb, reprint]{revtex4-1}

\usepackage[UKenglish]{babel}
\usepackage[latin9]{inputenc}
\usepackage[T1]{fontenc}
\usepackage{graphicx} \usepackage{graphics}
\usepackage{soul} 

	\newcommand{\Tc}{$T_{{\rm c}}$}
	\newcommand{\Jc}{$J_{{\rm c}}$}
	\newcommand{\Ba}{BaFe$_{1.8}$Co$_{0.2}$As$_2$}
	\newcommand{\Bax}{BaFe$_{2-x}$Co$_x$As$_2$}

\begin{document}

\author{S. Trommler} 
\altaffiliation{Dresden University of Technology, Department of Physics, Institute for Physics of Solids, 01062 Dresden, Germany}
\author{R. Hühne}
\author{J. Hänisch}
\author{E. Reich}
\author{K. Iida}
\author{S. Haindl}
\affiliation{IFW Dresden, P.O. Box 270116, 01171 Dresden, Germany, electronic mail: s.trommler@ifw-dresden.de}

\author{V. Matias}
\affiliation{Superconductivity Technology Center, Los Alamos National Laboratory, Los Alamos, New Mexico 87545; now at iBeam Materials, Santa Fe, New Mexico, electronic mail: vlado@iBeamMaterials.com}

\author{L. Schultz} 
\altaffiliation{Dresden University of Technology, Department of Physics, Institute for Physics of Solids, 01062 Dresden, Germany}
\author{B. Holzapfel  }
\altaffiliation{TU Bergakademie Freiberg,  09596 Freiberg, Germany }
\affiliation{IFW Dresden, P.O. Box 270116, 01171 Dresden, Germany, electronic mail: s.trommler@ifw-dresden.de}

\title{The influence of the buffer layer architecture on transport properties for \Ba\ films on technical substrates.}

\begin{abstract}

A low and almost temperature independent resistance in the normal state and an anomalous peak effect within the normal-superconducting transition have been observed in \Ba/Fe bilayers, prepared on IBAD-MgO/Y$_2$O$_3$ buffered technical substrates. 
A resistor network array sufficiently reproduces this effect, assuming an increase of the electrical conductance between tape and film with decreasing buffer layer thickness. Based on this model we evaluated the influence of this effect on the critical current density and successfully reconstructed the superconducting transition of the bilayer.

\end{abstract}

\maketitle


The development of the coated conductors technology proved high temperature superconductors to become commercially attractive in high current applications. Basically two approaches have been focused on: rolling assisted biaxially textured substrates (RABiTs) and ion beam assisted deposition (IBAD) of textured buffer layers on polycrystalline metal tapes\cite{Goy04, Iij04}. 
Recently, the growth of Fe-based superconductors such as \Bax\ (Ba-122) and FeSe$_{0.5}$Te$_{0.5}$ on IBAD-MgO/Y$_2$O$_3$/Hastelloy has been demonstrated. Their  high critical current densities at liquid He temperature as well as their high upper critical magnetic fields make them suitable for high-current applications \cite{ Iid11, Kat11, Li11}. 
In this letter we investigate the normal-superconducting transition for Ba-122 prepared on IBAD-MgO/Y$_2$O$_3$ buffered Hastelloy tapes with varying Y$_2$O$_3$ thickness. 
We will point out that the observed peak in the resistive transition can be modeled and quantitatively explained using a simple resistor network attributing good electrical conductance between Ba-122/Fe bilayer and tape in the case of a thinner buffer stack. Finally, we show that this model enables an accurate evaluation of important technical properties, such as critical temperature (\Tc) and critical current density (\Jc).

For comparison we used two different buffer architectures on polycrystalline C-276  tape (Hastelloy).
Tape (A) was planarized by solution deposition planarization (SDP) process using multiple chemical solution coatings of Y$_2$O$_3$ with a total thickness of approximately 1 $\mu$m.
To obtain a biaxially textured  buffer a MgO seed layer was prepared by IBAD at room temperature covered with homoepitaxial MgO of about 40 nm at 600 °C by electron beam sublimation \cite{She11}.
Tape (B) utilizes electro-polished Hastelloy, covered by a 10 nm thick amorphous electron-beam-deposited Y$_2$O$_3$ layer. On the 5 nm IBAD-MgO template 160 nm homoepitaxial MgO was deposited with electron beam sublimation\cite{Mat09}.
The thicker Y$_2$O$_3$ buffer is commonly used to prevent oxide superconductors from deteriorating the superconducting properties due to a reduced diffusion of transition metal ions from the tape into the film \cite{Mat10}.

\begin{figure}[ht]
\begin{center}\includegraphics[width=8cm]{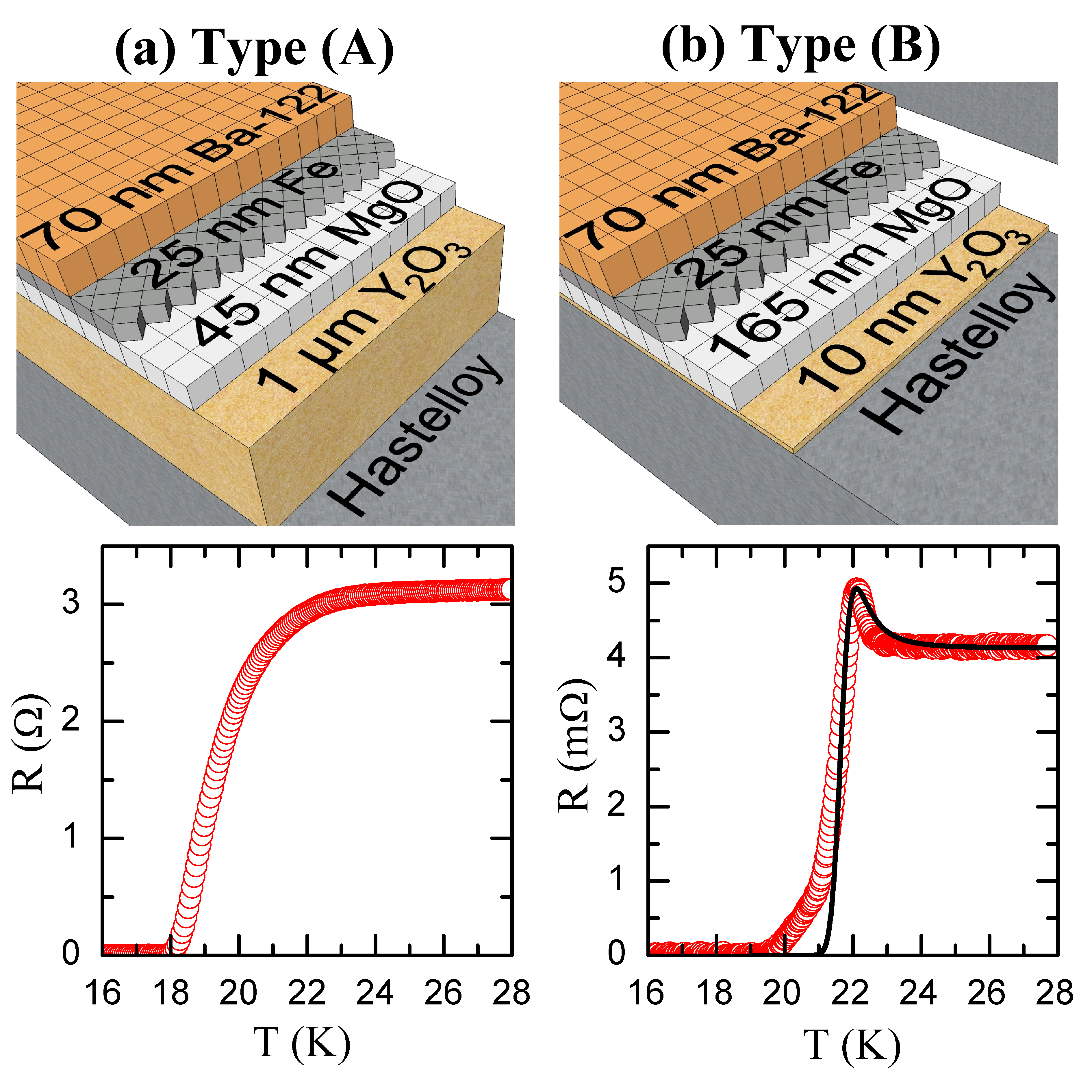}\end{center}
\caption{(a) Layer architecture for tape (A) and corresponding resistive superconducting transition; (b) layer architecture for tape (B) and corresponding superconducting transition. The solid line represents a fit according the introduced resistor network.} \label{figure_1}
\end{figure}

For both tapes, (A) and (B), a Ba-122/Fe bilayer was prepared in a pulsed laser deposition (PLD) setup with a base pressure of 10$^{-9}$ mbar applying in-situ reflection high energy diffraction (RHEED) to monitor the film evolution \cite{Iid11}.
70 nm Ba-122 were deposited on 25 nm Fe, which was shown to be beneficial for the textured growth of Ba-122 \cite{ The10}.
After cooling to room temperature the sample was covered with Au without breaking the vacuum to ensure low contact resistance. 
X-ray diffraction proves the phase formation of Ba-122, whereas pole figure measurements as well as RHEED  verify the epitaxial relationship  (001)[100]Ba-122 $\parallel $ (001)[110]Fe $\parallel $ (001)[100]MgO \cite{Iid11}. 
Transport measurements were performed in four-probe geometry using a Physical Property Measurement System (PPMS; Quantum Design). For the determination of \Jc\  tracks were prepared by ion etching with 1 mm width and 1 mm length.

The normal-superconducting transition of Ba-122 films on the tapes (A), displayed in fig.\ref{figure_1}(a), have a typical shape comparable to films prepared on single crystals  \cite{Iid10_2}. 
In contrast, films prepared on substrate type (B) (thin Y$_2$O$_3$ layer) exhibit a remarkable shape of the transition as illustrated in fig.\ref{figure_1}(b).
The resistance in the normal state is three orders of magnitude smaller with almost no temperature dependence. 
Close above the superconducting transition the resistivity peaks, followed by a sharp drop to zero. These features are common for films prepared on tapes (B) independently of \Tc\ and preparation conditions. Hence, they are a consequence of the tape architecture. Peak effects in the temperature dependent resistivity have also been observed in underdoped \Bax. However, they are accompanied by a reduced \Tc\ and can not explain our observations \cite{Ni08}. 

\begin{figure}[ht]
\begin{center}\includegraphics[width=8cm]{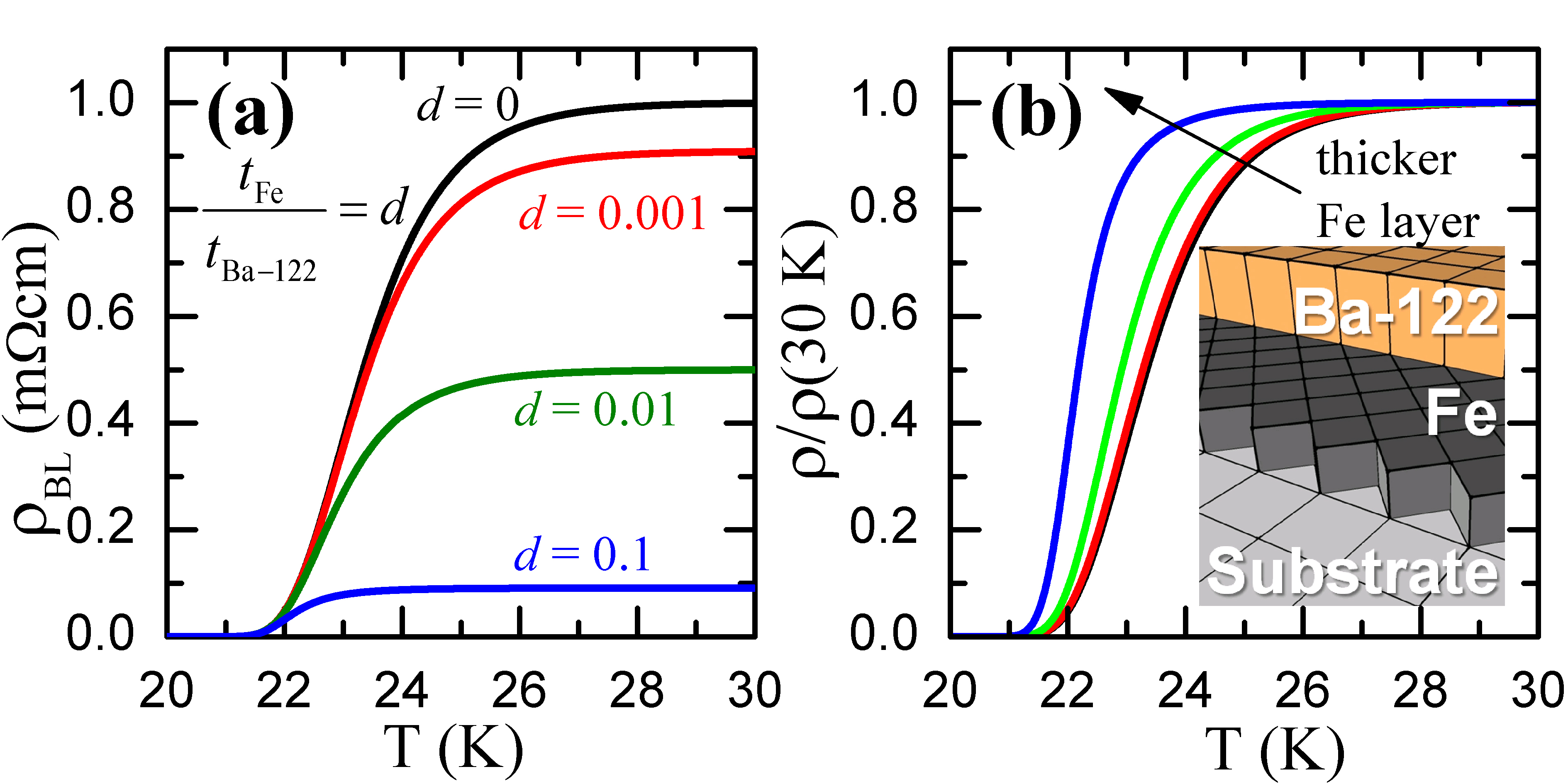}\end{center}
\caption{The simulated resistivity of the Ba-122/Fe bilayer ($\rho_\mathrm{BL}$) decreases with increasing Fe layer thickness $t_\mathrm{Fe}$ to Ba-122 layer thickness $t_\mathrm{Ba-122}$ ratio $d$ (a). As consequence the superconducting transition width reduces (b) and the T$_{c,90}$  criterion shifts to lower temperatures.} \label{figure_iron}
\end{figure}

The reduction of the normal state resistivity of Ba-122/Fe bilayers ($\rho_\mathrm{BL}$) with increasing ratio in thickness between Fe ($t_\mathrm{Fe}$) and Ba-122 ($t_\mathrm{Ba-122}$) was reported previously \cite{Iid10_2}. This effect is demonstrated in fig.\ref{figure_iron}(a), assuming a simple parallel circuit representing the resistivity of the Fe ($\rho_\mathrm{Fe}$) and the Ba-122 layer ($\rho_\mathrm{Ba-122}$), where $d=t_\mathrm{Fe}/t_\mathrm{Ba-122}$. The superconducting transition can be described as:

\begin{equation}
\rho_\mathrm{Ba-122}=\rho_\mathrm{n} \exp ((-T/T_\mathrm{c})^{-q})
\label{transition}
\end{equation}

where $\rho_\mathrm{n} \approx 1 $ m$ \Omega$cm corresponds to the Ba-122 resistivity in the normal state and was taken from single crystal data \cite{Ni08}.
The parameter $q$ determines the width of the transition and is set to 25 in our case. The resistivity of pure Fe amounts to 0.01 $\mu\Omega$cm at 20 K and increases about 1 $\mu\Omega$cm per 1 at\% Co diffusion \cite{Landolt}. Due to the unknown Co content of the Fe layer its resistivity $\rho_\mathrm{Fe}$ was estimated to a upper limit of 10 $\mu\Omega$cm.  
In general, both, $\rho_\mathrm{n}$ and $\rho_\mathrm{Fe}$, are temperature dependent functions, but were set constant in first-order approximation. 

The thickness of Fe in Ba-122/Fe bilayers has to exceed the threshold for a closed layer, which is between 4 nm and 15 nm, to ensure a complete coverage film for the epitaxial growth of Ba-122 \cite{Iid10_2}.
Hence, the resulting thickness ratios ($d$) are in the order of 0.1 including an important consequence, displayed in figure 2(b): The width of the superconducting transition of the Ba-122/Fe bilayer decreases with increasing Fe layer thickness. Therefore, its transition temperature is not identical to the single Ba-122 layer using the same criterion. 
It has to be considered that the evaluated \Tc $(B)$ from transport measurements of bilayers differs from the values of the bare superconducting compound, which is especially important regarding application of multilayered film architectures.

The description of the resistive transition for tape (B) needs additional considerations to be taken into account. 
A significant consequence of the varying buffer thickness is the influence on the electrical barrier properties between superconductor and tape. In the case of tape (B) an electrical connection can arise from defects in tape which give rise to pinholes in the buffer. This presumption is supported by the fact that in contrast to the used electro-polished tape comparable buffer stacks on smoother mechanically polished tapes are completely insulating. Additionally diffusion of metal ions from tape and Fe buffer can increase the conductivity of the MgO buffer.
A thick Y$_2$O$_3$ buffer suppresses the diffusion and overgrows the defects. 
In the following we will show that a change of the oxide buffer layer resistance by two orders of magnitude gives rise to the observed resistance peak at \Tc\ for tape (B).

\begin{figure}[ht]
\begin{center}\includegraphics[width=8cm]{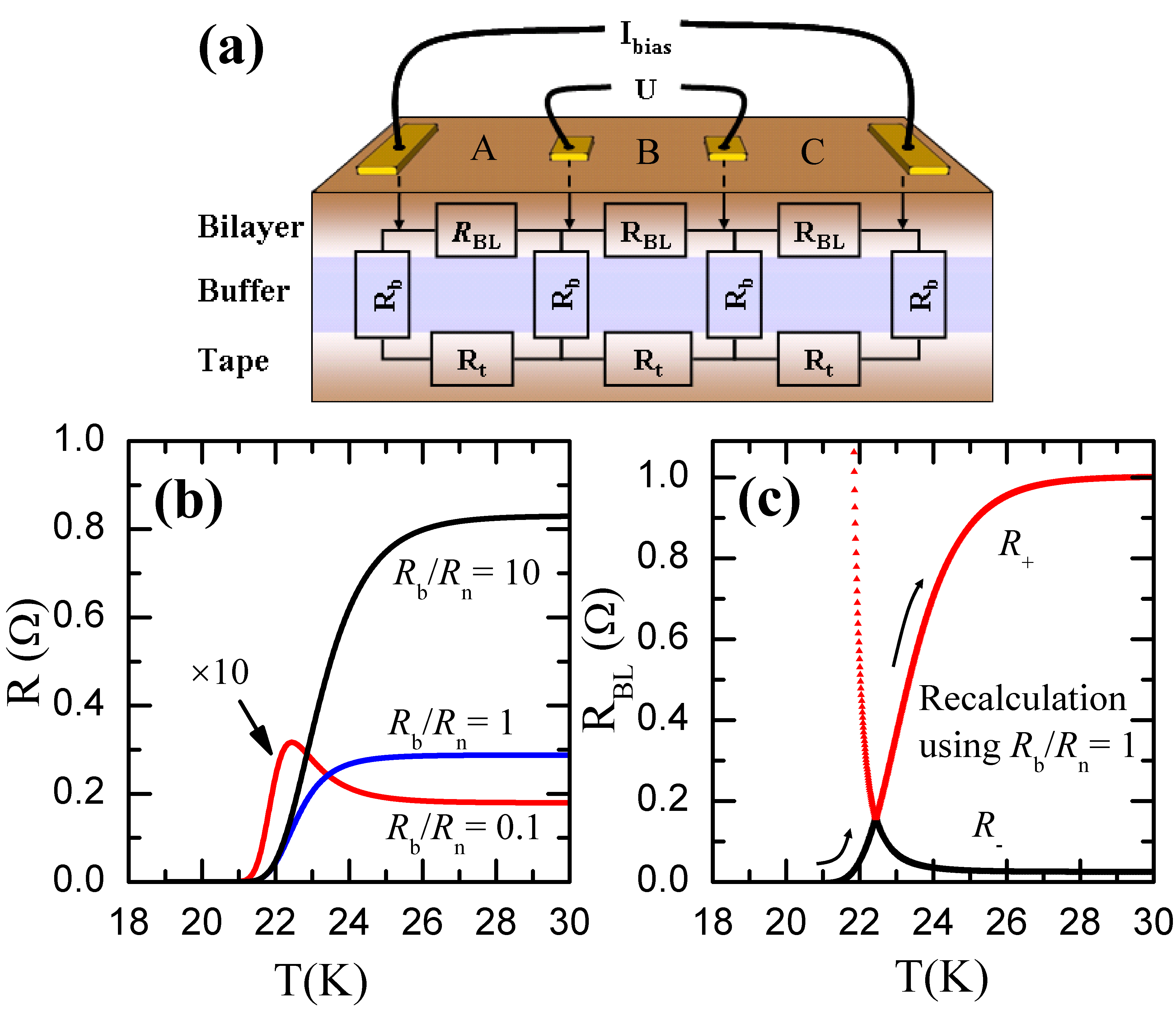}\end{center}
\caption{Schematic view of the four probe contact arrangement (a) including the assumed resistor network model. Simulation of the resistance, arising between the voltage leads for different barrier resistances (b) using eq. (\ref{R}) and ensuing solutions $R_{+}$ and $R_{-}$ for $R_\mathrm{BL}$ (c). The arrows indicate the resistive transition combining  the two solution branches.} 
\label{figure_2}
\end{figure}

Above \Tc\  the current is shunted by the tape, if the Ba-122 film and metal tape are electrically good connected,
as the resistivity of Ba-122 is one order of magnitude larger compared to Hastelloy \cite{Ni08,Lu08}. Therefore, most of the bias current flows through the tape, which gives only a small voltage drop between the voltage leads.
Within the superconducting transition the fraction of current through the film increases with decreasing resistance of the Ba-122 film resulting in an increasing voltage drop between the voltage leads, which finally reaches zero when the film becomes fully superconducting. 
Taking this into account, we split the sample geometry into three segments, separated by the voltage contacts, which is schematically illustrated in fig.\ref{figure_2}(a).
Based on these assumptions, a resistor network was modeled in which the conductance between tape and Ba-122/Fe bilayer is given by the resistivity of the MgO/Y$_2$O$_3$ buffer stack. 
For a homogeneous sample with constant width and an equidistant  lead arrangement we end up in the expression for the current $I$ passing the voltage leads: 

\begin{equation}
I = I_\mathrm{bias}\frac{2R_\mathrm{b}^2+4R_\mathrm{b}R_\mathrm{t}+R_\mathrm{t}(R_\mathrm{BL}+R_\mathrm{t})}{2R_\mathrm{b}^2+(R_\mathrm{BL}+R_\mathrm{t})^2+4R_\mathrm{b}(R_\mathrm{BL}+R_\mathrm{t})}
\label{R}
\end{equation}

where $I_\mathrm{bias}$ is the bias current and $R_\mathrm{BL}=\rho_\mathrm{BL}l/A$ is the resistance of Ba-122/Fe bilayer. $A$ represents the film cross section. The segments have a width $w=10$ mm and a length $l=2$ mm corresponding to the distance between the contacts.  $R_\mathrm{b}$ and $R_\mathrm{t}$ denote the resistance of buffer and tape, respectively. 
In general $R_\mathrm{b}$ and $R_\mathrm{t}$ are temperature dependent functions. However, we set them constant in first-order approximation due to the small temperature range which is treated. The resulting voltage drop between the voltage leads is given by $U=R_\mathrm{BL}I=RI_\mathrm{bias}$, where $R$ is the resultant measured resistance. 
The resistivity of polycrystalline Hastelloy exhibits only a small  temperature dependency and changes in total 3.5\% when cooled from room temperature to 10 K \cite{Lu08}.  If the current is shunted in the normal state, the transport properties are dominated by the tape, which explains the negligible temperature dependence of our data. 
With the resistivity of Hastelloy at 20 K, $\rho_\mathrm{t} $=1.23 $\mu \Omega$m, and a tape thickness of 70 $\mu$m we obtain $R_\mathrm{t} \approx  3.5$ m$\Omega $. 
The temperature dependent resistance of the Ba-122/Fe bilayer was combined to $R_\mathrm{BL}$ using a parallel circuit and eq.\ref{transition} as described previously. Its normal state resistance $R_\mathrm{n}$ was set to 1 $\Omega $, corresponding to typical values of Ba-122/Fe bilayers, prepared on MgO single crystals (MgO-SC).

Considering the difference of the buffer thickness for tape (A) and (B) the simulated resistive transition for decreasing $R_\mathrm{b}/R_\mathrm{n}$ is presented in fig.\ref{figure_2}(b). 
For the calculation of the resistive transition we set $T_c=23 $ K and $q=25$ in eq.(\ref{transition}). 
$R_\mathrm{b}/R_\mathrm{n}=10$ corresponds to a poor electrical connection between tape and superconducting film, corresponding to tape (A). With decreasing buffer thickness by one order of magnitude also $R_\mathrm{b}/R_\mathrm{n}$ decreases. Due to the reduced fraction of current flowing between the voltage leads for $R_\mathrm{b}/R_\mathrm{n}=1$ the normal state resistance is reduced, but no change in the transition shape occurs. A further decrease to $R_\mathrm{b}/R_\mathrm{n}=0.1$ corresponds to a good electrical connection between tape and superconducting film equivalent to tape (B). In this case the normal state resistance is drastically reduced and a peak appears next to the normal-superconducting transition in agreement with the observed behavior. This calculation confirms that a change in the electrical connection between tape and Ba-122/Fe bilayer gives rise to the described behavior.

For $R_\mathrm{b}/R_\mathrm{n} \rightarrow   0$ the normal state resistance converges to $R_\mathrm{t}$ and the peak in resistance vanishes as the resistor network collapses to a simple parallel circuit. This case corresponds to a direct connection of the superconducting film with a conducting substrate and was observed previously for Ba-122 films prepared on single crystalline SrTiO$_3$ substrates \cite{Iid09}. The substrate becomes conducting due to the loss of oxygen in SrTiO$_3$ at high temperatures in high vacuum,   which results in a three orders of magnitude lower normal state resistivity and a reduced width of the superconducting transition compared to films prepared on non-conducting oxide substrates. This estimation suggests that the observed peak arises only in a small range of the electrical barrier resistance between Ba-122/Fe bilayer and tape. An expansion of the model with additional resistor loops for the simulation of a continuous shunting ob bilayer and tape within the zones A, B and C does not affect the results above.

\begin{figure}[ht]
\begin{center}\includegraphics[width=8cm]{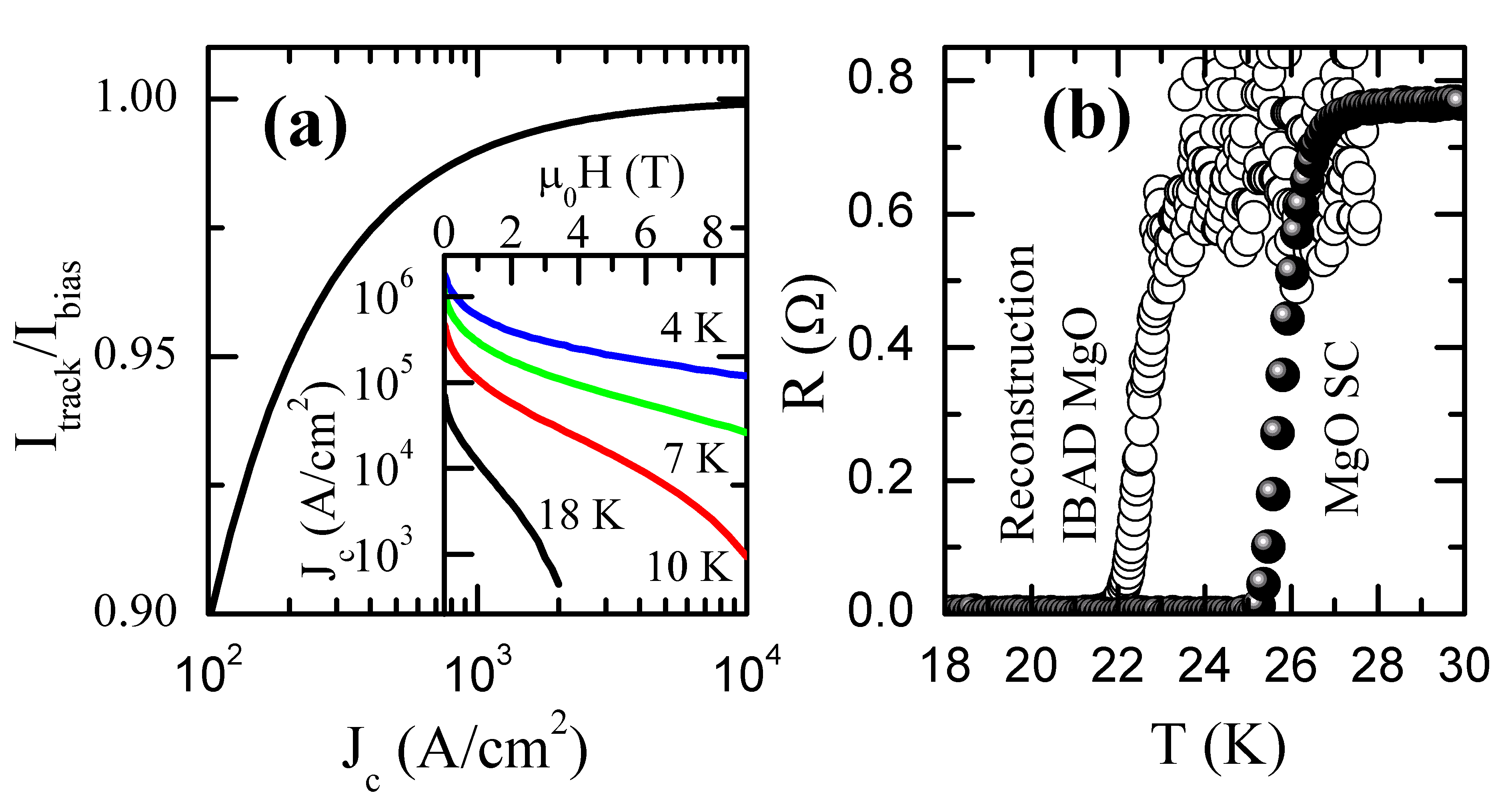}\end{center}
\caption{The calculated current which is shunted by the tape depending on the critical current density, determined at $E_c=1\mu$ V/cm (a) and measured field dependent critical current density of the sample (inset). Reconstructed transition of the Ba-122/Fe bilayer from measurement data of tape (B)(fig.\ref{figure_1}(b)) using the resistor network in comparison to a film prepared on a MgO single crystal (MgO-SC).} \label{Jc}
\end{figure}

In the following section we address the accurate estimation of the superconducting properties of the Ba-122/Fe bilayer for tape (B).
Solving the quadratic equation given by eq.(\ref{R}) results in two solutions for $R_\mathrm{BL}$ denoted as $R_{+}$ and $R_{-}$, which have to be combined in order to obtain the complete transition. This is illustrated   in fig.\ref{figure_2}(c) for $R_\mathrm{b}/R_\mathrm{n}=0.1$. The arrows indicate the resistive transition combining both solutions.

For the reconstruction of the transition from experimental data reasonable resistance values for buffer and tape have to be evaluated. 
In the normal state ($T=30$ K) the resistance of the bilayer is larger than that of tape and MgO/Y$_2$O$_3$ barrier. Therefore, we can assume $R_\mathrm{BL}\gg R_\mathrm{b},R_\mathrm{t}$ which gives $R_\mathrm{t}\sim R(30$ K$) = 4$ m$\Omega $ (fig.\ref{figure_iron}(b)). This value is in good agreement with the estimated $R_\mathrm{t}$ using the resistivity of Hastelloy.

Since the combined resistive transition $R_\mathrm{BL}$ has to be continuous, $R_{+}$ and $R_{-}$  have to be equal at their extremum which gives $R_\mathrm{b} \approx$ 7.8 $\mathrm{m}\Omega$. 
The reconstructed superconducting transition using these values is displayed in fig.\ref{Jc}(b) and compared to a Ba-122/Fe bilayer on MgO-SC prepared under comparable conditions. The \Tc, defined as 50\% of the normal resistivity at 30 K, was evaluated to 23 K which is slightly lower compared to films prepared on MgO-SC with \Tc $=26$ K. 
The origin of this deviation is not uniquely assignable. Both, a reduced preparation temperature while using metal substrates and additional doping of Ba-122 by  transition metal ions will affect \Tc.
The simulation of $R$ is in very good agreement with the observed transition of tape (B) using eq.(\ref{R}) with a Ba-122/Fe normal state resistance of 0.7 $\Omega $ and the determined values for $R_\mathrm{b}$ and $R_\mathrm{t}$  as displayed in fig.\ref{figure_1}(b).

Finally, we etched a 1 mm wide and 1 mm long track in section B (fig.\ref{figure_2}(a)) for the determination of  \Jc\ .
We made the following assumption in order to estimate the influence of this geometry on the fraction of current flowing through the tape at the determined \Jc\ of the track: 
Currents for which the track enters the flux flow state are still loss free in the remaining part of the Ba-122 section A and C due their larger width. Therefore one can reduce the resistor network to a simple parallel circuit, where the current  through the track ($I_\mathrm{track}$) is defined as $I_\mathrm{bias}=I_\mathrm{track}+I_\mathrm{tape}$  and the fraction shunted by the tape is $I_\mathrm{tape}=U_\mathrm{c}/(R_\mathrm{t}+2R_\mathrm{b}$). The field criterion $E_c=U_c/l$ with the length of the track $l=1$ mm was set to 10$^{-6}$ V/cm for the determination of \Jc.
The resulting ratio of $I_\mathrm{track}/I_\mathrm{bias}$ is given in fig.\ref{Jc}(a), where for \Jc\ higher than 1 kA/cm$^2$ 99\%  of the current passes the superconducting track. For the self field \Jc\ of the film, which is higher than 70 kA/cm up to $T=18$ K (fig.\ref{Jc}(a), inset), this correction is negligible.

In summary we have demonstrated that a peak in the resistive transition appears when the electrical barrier resistance between tape and Ba-122/Fe bilayer is reduced. A simple resistor network array describes all main features of the measured behavior. The error in the determination of the self field \Jc\ is negligible for values higher than 1 kA/cm$^2$.
The results indicate that a reduced buffer layer thickness has no detrimental effect on the superconducting properties of Ba-122. A thin electrical conducting barrier layer may even be beneficial as it provides an effective protection of the superconducting film in case \Jc\ is exceeded, as the majority of the current will be shunted by the tape.

This work was partially supported by the Deutsche Forschungsgesellschaft and SuperIron, founded by the European Commission.

\bibliographystyle{apsrev}

\end{document}